\documentclass[9pt,conference]{IEEEtran}
\IEEEoverridecommandlockouts

\usepackage{cite}
\usepackage{amsmath,amssymb,amsfonts}
\usepackage{algorithmic}
\usepackage{graphicx}
\usepackage{textcomp}
\usepackage{xcolor}
\usepackage{hyperref}
\usepackage{multirow}
\usepackage{booktabs}
\usepackage[misc]{ifsym}
\def\BibTeX{{\rm B\kern-.05em{\sc i\kern-.025em b}\kern-.08em
    T\kern-.1667em\lower.7ex\hbox{E}\kern-.125emX}}
\begin{document}
\title{CAMEL: Cross-Attention Enhanced Mixture-of-Experts and Language Bias for Code-Switching Speech Recognition}
\author{
    \IEEEauthorblockN{
        He Wang\IEEEauthorrefmark{1}, 
        Xucheng Wan\IEEEauthorrefmark{2}, 
        Naijun Zheng\IEEEauthorrefmark{2}, 
        Kai Liu\IEEEauthorrefmark{2}, 
        Huan Zhou\IEEEauthorrefmark{2}, 
        Guojian Li\IEEEauthorrefmark{1}
        Lei Xie\IEEEauthorrefmark{1}$^\text{\Letter}$ \thanks{\Letter~Lei Xie is the corresponding author.}
  }
    \IEEEauthorblockA{
        \IEEEauthorrefmark{1}Audio, Speech and Language Processing Group (ASLP@NPU), School of Computer Science\\ Northwestern Polytechnical University, Xi'an, China\\
        Email: \{hwang2001,aslp\_lgj\}@mail.nwpu.edu.cn, lxie@nwpu.edu.cn
    }
    \IEEEauthorblockA{
        \IEEEauthorrefmark{2}IT Innovation and Research Center \\ Huawei Technologies, Shenzhen, China\\
        Email: \{wanxucheng,zhengnaijun,liukai89,zhou.huan\}@huawei.com
    }
}


\maketitle

\begin{abstract}

Code-switching automatic speech recognition (ASR) aims to transcribe speech that contains two or more languages accurately.
To better capture language-specific speech representations and address language confusion in code-switching ASR, the mixture-of-experts (MoE) architecture and an additional language diarization (LD) decoder are commonly employed. 
However, most researches remain stagnant in simple operations like weighted summation or concatenation to fuse language-specific speech representations, leaving significant opportunities to explore the enhancement of integrating language bias information.
In this paper, we introduce CAMEL, a cross-attention-based MoE and language bias approach for code-switching ASR. 
Specifically, after each MoE layer, we fuse language-specific speech representations with cross-attention, leveraging its strong contextual modeling abilities. 
Additionally, we design a source attention-based mechanism to incorporate the language information from the LD decoder output into text embeddings. 
Experimental results demonstrate that our approach achieves state-of-the-art performance on the SEAME, ASRU200, and ASRU700+LibriSpeech460 Mandarin-English code-switching ASR datasets.

\end{abstract}

\begin{IEEEkeywords}
code-switching, speech recognition, cross-attention, mixture-of-experts, language bias
\end{IEEEkeywords}

\section{Introduction}
With the advancement of deep learning, automatic speech recognition (ASR) has made significant strides, surpassing human-level performance on some open-source benchmarks and achieving excellent results in monolingual speech recognition tasks~\cite{xiong2017toward}. 
However, code-switching ASR remains a challenge. 
This is mainly due to the lack of manually annotated code-switching data and also due to the auditory similarities between different languages caused by accents or tones, known as language confusion. 
To address the data scarcity issue, methods such as generating code-switching text and speech data~\cite{du2021data,yu2023code,hu2023improving,hussein2024speech} and utilizing large-scale unlabeled data for unsupervised learning~\cite{ogunremi2023multilingual,abdallah2024leveraging} have been proposed, yielding some benefits for code-switching ASR.
To mitigate language confusion, current research mainly focuses on two aspects. 
The first is at the encoder level, where the encoder architecture is well-designed to efficiently model language-specific acoustic representations. 
The second approach involves introducing additional classifiers or decoders to model frame-level or token-level language ID (LID) or language boundaries, which incorporates more language information into the model.

First, regarding the modeling of language-specific acoustic representations: Bi-encoder based mixture-of-experts (MoE) architectures~\cite{lu2020bi,song2022language,yang2024effective,wang2024tri} for code-switching ASR employ two separate encoders as language experts, thereby capturing language-specific acoustic features.
However, they can hardly learn any cross-lingual information due to the fully decoupled model structure. 
The LAE~\cite{tian2022lae} alleviates this issue by introducing a shared encoder before the language-specific encoders to model speech in different languages.
Unlike the above approaches, which use encoders as language experts, the BA-MoE~\cite{chen2023ba} proposes the MoE-Adapter module. 
This module aims to use a lightweight structure, such as a feed-forward network, as language experts embedded within each encoder layer, while a linear gating unit performs a frame-level weighted summation of the two language speech features. 
Despite the great improvement of BA-MoE in code-switching ASR, the linear gate fusion method is ineffective at capturing cross-lingual contextual information. 
Meanwhile, cross-attention~\cite{guo2023npu,zhang2023ve,wang2024mlca} has shown powerful capability in this regard.

Second, regarding the incorporation of language information: BA-MoE~\cite{chen2023ba} employs an additional decoder to capture the position where code-switching occurs, named language boundary. 
Language alignment loss (LAL)~\cite{liu2024aligning} is proposed to enrich bilingual ASR models with language information, including language classification and code-switching boundary.
Aditya et al.~\cite{aditya2024attention} propose an attention-guided loss that activates language attention heads in the Whisper~\cite{radford2023robust} decoder through adapter training, thereby improving Whisper in recognizing code-switching speech. 
Liu et al.~\cite{liu2023reducing,liu2024enhancing}  propose a series of approaches combining an additional language diarization (LD) decoder with language posterior bias. 
This method uses the LD decoder to obtain token-level language classification posteriors, concatenated with text embeddings, and then fed into the main decoder.
However, directly concatenating low-dimension classification posteriors with text embeddings makes it challenging for the model to effectively focus on the language information.

Inspired by the above two aspects for improving code-switching ASR, we further explore the fusion method of different language speech representations and the effective way to incorporate language information.
In this paper, we propose the \textbf{C}ross-\textbf{A}ttention enhanced \textbf{M}o\textbf{E} and \textbf{L}anguage bias (CAMEL) approach for code-switching ASR.
Specifically, we use E-Branchformer~\cite{kim2023branchformer} as the backbone, with the encoder comprising Branchformer blocks and additional MoE layers.
After each MoE layer, an MoE-Adapter is utilized to perform adaptation learning for different languages, followed by the fusion of language-specific representations using gated cross-attention. 
Additionally, language-wise connectionist temporal classification (CTC)~\cite{graves2012connectionist} is introduced to guide the MoE-Adapter in learning the speech features of the specific language.
Moreover, we introduce an LD decoder and use cross-attention to incorporate the LD decoder output into the text embeddings for language bias, enriching the system with language information. 
Finally, experimental results demonstrate that our proposed approach achieves state-of-the-art (SOTA) performance on major Mandarin-English code-switching datasets, including SEAME~\cite{lyu2010seame}, ASRU200~\cite{shi2020asru}, and ASRU700+LibriSpeech460~\cite{shi2020asru,panayotov2015librispeech}, showing a significant advantage over existing code-switching ASR approaches.
\begin{figure}[tbp]
\centerline{\includegraphics[width=1.0\columnwidth]{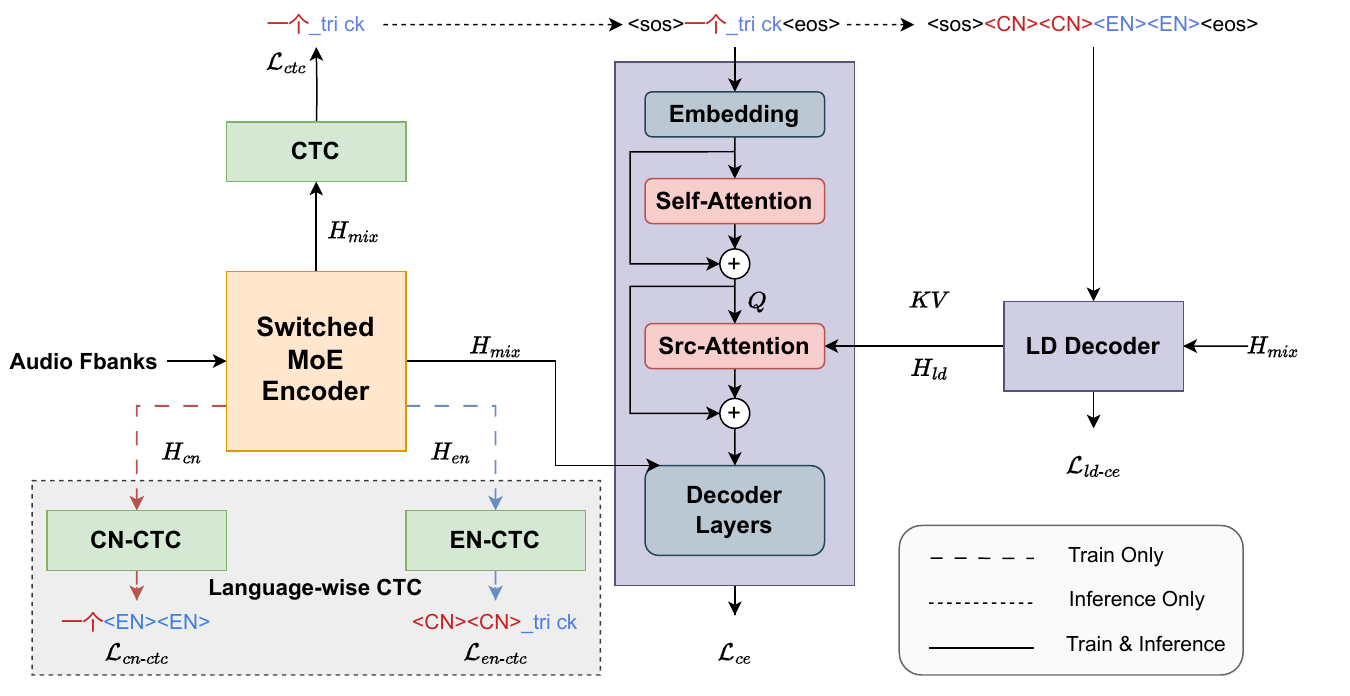}}
\caption{Overall system architecture of our proposed approach. The English translation of input utterance is "One Trick".}
\label{fig-overall}
\end{figure}
\section{Method}
\begin{figure*}[tbp]
\centerline{\includegraphics[width=2.0\columnwidth]{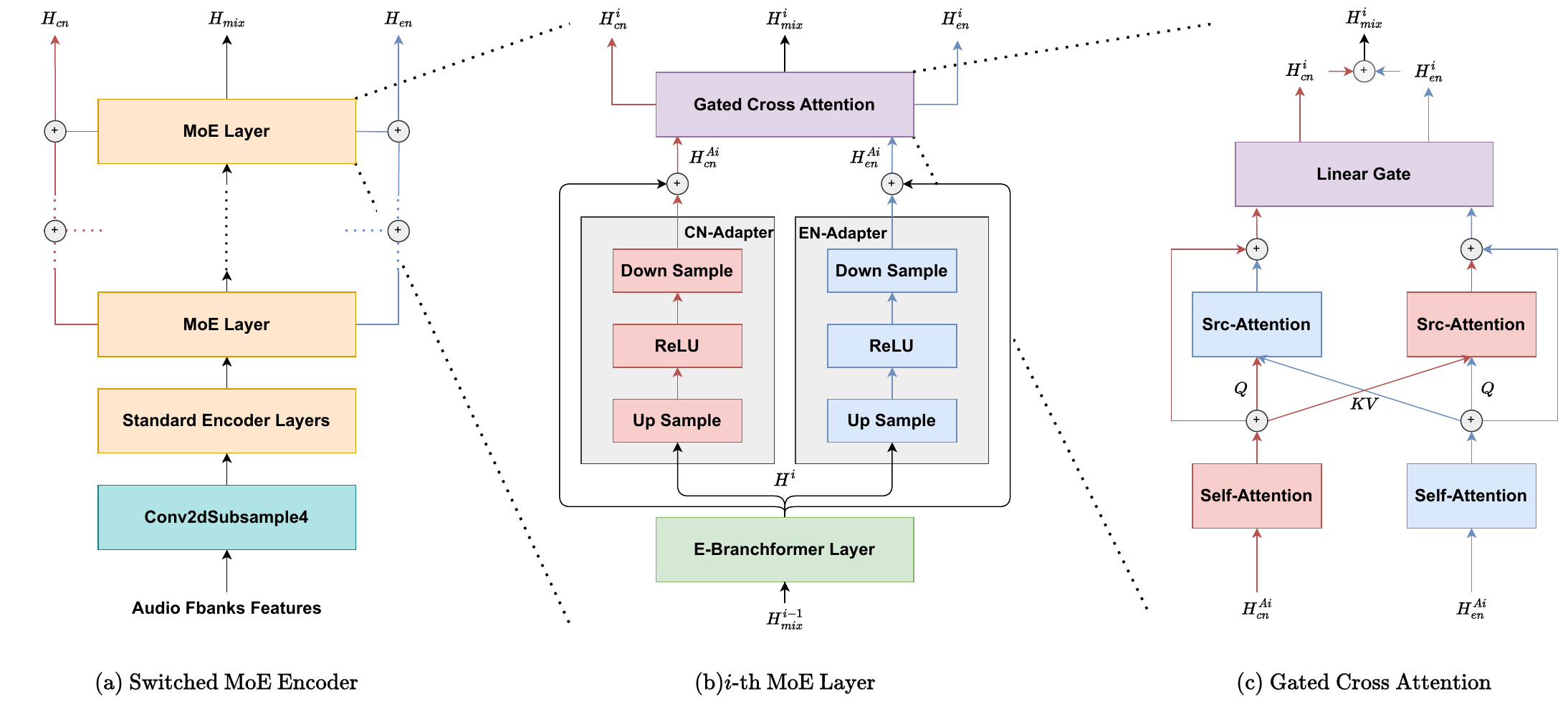}}
\caption{(a) The overall structure of switched MoE Encoder, (b) MoE Layer architecture, (c) Detailed design of gated cross-attention.}
\label{fig-encoder}
\end{figure*}

Our proposed CAMEL code-switching ASR system primarily consists of four modules: encoder, main decoder, LD decoder, and CTC. 
The overall system structure is shown in Figure~\ref{fig-overall}. 
Specifically, we use E-Branchformer as the backbone to build the switched MoE encoder~\cite{ye2024sc}, comprising standard encoder layers and MoE layers, as shown in Figure~\ref{fig-encoder}(a). 
The input audio Fbank features are firstly downsampled with a factor of 4 by 2-dimension convolution, and then modeled by standard encoder layers to get stable speech representations and then fed into MoE layers. 
At each MoE layer, we use MoE-Adapter, including CN-Adapter and EN-Adapter, to adaptively extract language-specific speech representations from the encoder layer's output, and fuse the representations with a gated cross-attention module,  as shown in Figure~\ref{fig-encoder}(b) and \ref{fig-encoder}(c).
The above adaptation learning and fusion process is guided by the language-wise CTC loss.
The main decoder and LD decoder share almost same model structure.
The difference lies in the fact that the main decoder has an attention module to incorporate the language information from LD decoder output.
Moreover, the main decoder receives text as input while the LD decoder receives language sequence labels as input.
In the main decoder, text labels are passed through an embedding layer to obtain text embeddings, which serve as the query, while the output of the LD decoder acts as the key and value. 
The attention module, similar to a Transformer decoder layer, is then employed to bias text embeddings with the LD decoder output.
\subsection{Language Diarization Decoder for Language Bias}
In the CAMEL system, we employ an additional Transformer-based decoder to model the token-level language information of the text sequence, referred to as the language diarization (LD) decoder, as shown in Figure~\ref{fig-overall}. 
During training, the LD decoder takes the ground-truth language label sequence and the encoder output as inputs, producing the output \( H_{ld} \).
It is then used as the key ($K$) and value ($V$), while the decoder Embedding output is first modeled by multi-head self-attention with residual connection, then the output serves as the query ($Q$) in the multi-head source attention (src-attention) module.
Both self-attention and src-attention follow the dot-product attention mechanism~\cite{vaswani2017attention}.
The output of src-attention with residual connection is then fed into the Transformer decoder layers for further modeling, achieving the incorporation of language information into the text embedding, also known as language bias.
During inference, we use the attention-rescoring strategy. 
First, we generate several candidate hypotheses and their CTC scores using CTC prefix beam-search.
Then, these hypotheses and their corresponding language label sequences are input into the decoder and LD decoder, respectively.
The decoder outputs an attention score for each candidate transcript. 
Finally, the hypothesis with the highest combined score from both CTC and attention is selected as the final decoding result.
\subsection{MoE layer with Gated Cross Attention}
To better model language-specific speech representations while preserving cross-lingual contextual information, our approach avoids using dedicated encoders as language experts, as in the Bi-encoder models. 
Instead, we use lightweight adapter modules to adapt the speech representations output by the E-Branchformer layer to the corresponding language space.
Compared to BA-MoE~\cite{chen2023ba}, we focus more on modeling cross-lingual contextual information and design a gated cross-attention module in the MoE part. We enhance the linear gating unit by adding a cross-attention module, which is effective in modeling contextual relationships between different speech representations, and illustration of this module is shown in Figure~\ref{fig-encoder}(c). 

The MoE layer consists of a standard E-Branchformer layer, MoE-Adapter, and gated cross-attention. 
Specifically, the MoE-Adapter comprises CN-Adapter and EN-Adapter modules, each consisting of a linear up-sampling layer, a ReLU layer, and a linear down-sampling layer. 
The adapted output is then residual-connected with the output of the E-Branchformer layer.
Considering the \(i\)th MoE layer, its input \(H_{mix}^{i-1}\) is the output from the previous encoder layer.
First, it goes through an E-Branchformer layer and comes to $H^{i}$.
The process of the MoE-Adapter can be described by the following formula:
\begin{equation}
    \begin{aligned}
        & H_{en}^{Ai}=H^{i}+W_{en}^{D}(\text{ReLU}(W_{en}^{U}(\text{LayerNorm}(H^{i})))), \\
        & H_{cn}^{Ai}=H^{i}+W_{cn}^{D}(\text{ReLU}(W_{cn}^{U}(\text{LayerNorm}(H^{i})))),
    \end{aligned}
\end{equation}
Where $W_{en}^{U}$ and $W_{en}^{D}$ are the weight matrices for the linear up-sampling and down-sampling in the EN-Adapter, respectively, and $W_{cn}^{U}$ and $W_{cn}^{D}$ are the weight matrices for those in the CN-Adapter. 
$H_{en}^{A_i}$ and $H_{cn}^{A_i}$ represent the English and Mandarin representations obtained after adaptation, respectively.

The gated cross-attention consists of a cross-attention and a linear gating unit.
Specifically, the cross-attention, which has a symmetric structure, is composed of multi-head self-attention and multi-head src-attention.
It takes the Mandarin and English speech representations output by the MoE-Adapter as inputs. 
Each representation first passes through multi-head self-attention to model global context within the self-language space.
Then, each representation is used as the query, while the representation of the other language serves as the key and value for multi-head src-attention.
Through this, the cross-lingual contextual relationship between the two language speech representations can be well modeled.
Residual connection follows both the multi-head self-attention and multi-head src-attention.
The computation process of cross-attention can be represented as follows:
\begin{equation}
    \begin{aligned}
        & H_{en}^{Self_i} = H_{en}^{A_i} + \text{MHSelfAttention}(H_{en}^{A_i}), \\
        & H_{cn}^{Self_i} = H_{cn}^{A_i} + \text{MHSelfAttention}(H_{cn}^{A_i}),
    \end{aligned}
\end{equation}
\begin{equation}
    \begin{aligned}
    & H_{en}^{Self_i}\xrightarrow{\text{linear}}Q_{en}; H_{cn}^{Self_i}\xrightarrow{\text{linear}}K_{cn}, V_{cn},\\
    & H_{cn}^{Self_i}\xrightarrow{\text{linear}}Q_{cn}; H_{en}^{Self_i}\xrightarrow{\text{linear}}K_{en}, V_{en},
    \end{aligned}
\end{equation}
\begin{equation}
    \begin{aligned}
        & H_{en}^{Src_i} = H_{en}^{Self_i} + \text{MHSrcAttention}(Q_{en}, K_{cn}, V_{cn}), \\
        & H_{cn}^{Src_i} = H_{cn}^{Self_i} + \text{MHSrcAttention}(Q_{cn}, K_{en}, V_{en}),
    \end{aligned}
\end{equation}
where \(\text{MHSelfAttention}\) and \(\text{MHSrcAttention}\) represent the multi-head self-attention and multi-head src-attention, respectively. 
$H_{en}^{Self_i}$ and $H_{cn}^{Self_i}$ denote the English and Mandarin speech representations after multi-head self-attention and residual connection, while $H_{en}^{Src_i}$ and $H_{cn}^{Src_i}$ represent those after multi-head src-attention and residual connection.
The symbol \(\xrightarrow{\text{linear}}\) indicates a linear projection.

The linear gating unit consists of one linear projection layer and one softmax function and it performs weighted summation for the two language speech representations obtained from the dual attention paths.
Considering that they both have $T$ feature frames and $F$ feature channels, this process can be formulated as:
\begin{equation}
    \begin{aligned}
        [w_{en}^{i}, w_{cn}^{i}] = & \text{softmax}(H_{en}^{Src_i}W^G + H_{cn}^{Src_i}W^G), \\
        H_{en}^{i} = & w_{en}^{i} \cdot H_{en}^{Src_i}; H_{cn}^{i} = w_{cn}^{i} \cdot H_{cn}^{Src_i}, \\
        & H_{mix}^{i} = H_{en}^{i} + H_{cn}^{i},
    \end{aligned}
\end{equation}
where $W^G\in\mathbb{R}^{F\times2}$ is the weight matrix for the linear projection. $w_{en}^{i}\in\mathbb{R}^{T\times1}$ and $w_{cn}^{i}\in\mathbb{R}^{T\times1}$ represent the weights for English and Mandarin speech representations.


\subsection{Loss Function}
We design a language-wise CTC loss function to guide the MoE-Adapter and gated cross-attention in extracting language-specific speech representations and modeling cross-lingual contextual information.
To be specific, we average the English and Mandarin speech representation output from each MoE layer to obtain $H_{en}$ and \( H_{cn} \). 
Given that the switched MoE encoder has $L$ layers in total, with $L/2$ being MoE layers, the formula is as follows:
\begin{equation}
    H_{en}=\frac{1}{2L}\sum_{i=1}^{L/2}{H_{en}^{i}};H_{cn}=\frac{1}{2L}\sum_{i=1}^{L/2}{H_{cn}^{i}},
\end{equation}
As shown in Figure~\ref{fig-overall}, the English and Mandarin speech representations are paired with corresponding EN-CTC and CN-CTC to calculate the CTC loss for each language, named language-wise CTC. 
It is important to note that for EN-CTC, the target label sequence is derived from the original text sequence by replacing all Mandarin tokens with $\textless\text{CN}\textgreater$. 
Similarly, for CN-CTC, it is obtained by replacing all English tokens with the $\textless\text{EN}\textgreater$. 
Besides, these $\textless\text{EN}\textgreater$ and $\textless\text{CN}\textgreater$ tokens share the same token ID as the language sequence tokens input into the LD decoder.

The overall loss function of the CAMEL system is composed of CTC loss and cross-entropy (CE) losses.
The CTC loss of the system consists of three parts: the CTC losses calculated from the encoder output $L_{ctc}$, from English speech representation $L_{en\text{-}ctc}$, and from Mandarin speech representation $L_{cn\text{-}ctc}$.
The CE loss comprises two parts: \(\mathcal{L}_{ce}\), which is the CE loss calculated between the predicted text sequence from the decoder and the ground-truth text sequence, and \(\mathcal{L}_{ld\text{-}ce}\), which is calculated between the predicted language sequence from the LD decoder and the ground-truth.
The overall loss function for the model is as follows:
\begin{equation}
\mathcal{L}_{lang\text{-}ctc} = \frac{\mathcal{L}_{en\text{-}ctc}+\mathcal{L}_{cn\text{-}ctc}}{2},
\end{equation}
\begin{equation}
\mathcal{L}=\lambda(\alpha\mathcal{L}_{lang\text{-}ctc}+(1-\alpha)\mathcal{L}_{ctc}) + (1-\lambda)\mathcal{L}_{ce} + \beta\mathcal{L}_{ld\text{-}ce},
\end{equation}
where, \(\lambda\), \(\alpha\), and \(\beta\) are weighting factors for different loss components, set to 0.3, 0.3, and 0.8, respectively, in this work.

\section{Experiments}
\subsection{Datasets}
In this paper, we conduct experiments on three common Mandarin-English code-switching datasets to show the effectiveness of our proposed CAMEL system. 
The details of these datasets are as follows.\\
\textbf{SEAME}. A spontaneous code-switching corpus recorded by Southeast Asian speakers. 
Both intra- and intersentence code-switching speech exist in this dataset.
The total duration of recorded audio is about 115 hours, including a training set and two test sets named as Dev\_MAN and Dev\_SEG, in the same manner as \cite{liu2023reducing,yang2024effective,liu2024aligning}.\\
\textbf{ASRU700+LibriSpeech460}. ASRU-2019 Mandarin-English code-switching challenge dataset, which has about 500h Mandarin training set, 200h Mandarin-English code-switching training set, and a 20h test set, named ASRU\_Test, along with the train-clean subset of 460 hours from Librispeech, consistent with \cite{chen2023ba,ye2024sc}. \\
\textbf{ASRU200}. Only use the 200h code-switching subset of the ASRU-2019 Mandarin-English code-switching challenge dataset as the training set, and ASRU\_Test as the test set, consistent with \cite{liu2024enhancing,liu2024aligning}.
\subsection{Setup}
All models are built using the WeNet open-source toolkit~\cite{yao2021wenet}.
Experiments on these three datasets all follow the same model configurations to build the baseline and conduct ablation studies. 

The \textbf{Baseline} model employs a 12-layer standard E-Branchformer encoder, with 4 attention heads of dimension 256 and one feed-forward unit of size 1024. 
The decoder uses a 6-layer Transformer decoder.
The \textbf{S1} system extends the baseline by adding MoE-Adapters to the end of each of the last 6 encoder layers, and it is trained using language-wise CTC.
The \textbf{S2} system is built on S1 by adding a linear gating unit after all MoE-Adapters.
The \textbf{S3} system further enhances S2 by replacing the linear gating unit with the gated cross-attention module to improve cross-lingual contextual modeling. 
Additionally, in the 6 MoE layers, the gated cross-attention modules share weights every 2 layers.
The \textbf{CAMEL} system, our proposed, extends S3 by adding an LD decoder, incorporating the language-informed output into the decoder's text embeddings through attention.

All experiments use a Mandarin character + English Byte Pair Encoding (BPE) modeling approach for the dictionary, with 3000 BPE tokens for the SEAME and ASRU200 datasets, and 4500 BPE tokens for the ASRU700+LibriSpeech460 dataset. 
Moreover, all models are trained for 150 epochs with Adam optimizer and learning rate of 0.001.
After training, we average 30 epoch models with the lowest loss on the development set for inference.
\subsection{Main Results}
\begin{table}[t]
    \centering
    \caption{Results on the SEAME dataset in terms of English word error rate (WER), Mandarin character error rate (CER), and Mandarin-English mixed error rate (MER) (\%):}
    \resizebox{\linewidth}{!}{
        \begin{tabular}{cc|ccc|ccc}
            \toprule
            \multirow{2}{*}{System} & \multirow{2}{*}{Params.} & \multicolumn{3}{c|}{Dev\_MAN} & \multicolumn{3}{c}{Dev\_SEG}  \\
             \cline{3-5} \cline{6-8} ~ & ~ & MER & CER & WER & MER & CER & WER \\
            \hline
            MoE-LAE~\cite{yang2024effective} & 24.5M & 20.7 & - & - & 29.0 & - & - \\
            LAL~\cite{liu2024aligning} & 48.3M & 16.4 & 14.8 & 29.1 & 23.2 & 21.7 & 28.3 \\
            LPB~\cite{liu2023reducing} & - & 16.3 & - & - & 23.0 & - & - \\
            \midrule
            Baseline & 39.9M & 16.13 & 13.14 & 24.92 & 22.89 & 16.74 & 26.59 \\
            S1 & 40.3M & 16.02 & 13.07 & 24.68 & 22.83 & 16.87 & 26.41 \\
            S2 & 40.3M & 15.86 & 13.02 & 24.22 & 22.60 & 16.59 & 26.22 \\
            S3 & 46.3M & 15.56 & 12.79 & 23.69 & 22.14 & 16.02 & 25.83 \\
            CAMEL & 55.3M & \textbf{15.32} & 12.48 & 23.66 & \textbf{21.84} & 15.92 & 25.40 \\
            \bottomrule
        \end{tabular}
    }
    \label{table-seame}
\end{table}
\subsubsection{SEAME}
Table~\ref{table-seame} shows the experimental results on the SEAME dataset for the E-Branchformer baseline, our proposed CAMEL system, and other recent SOTA methods.
Compared to the baseline, the CAMEL system achieved a relative reduction of 5\% (16.13\% → 15.32\%) in MER on the Dev\_MAN test set and 4.6\% (22.89\% → 21.84\%) on the Dev\_SEG test set. 
This demonstrates that CAMEL significantly improves performance on both Mandarin dominant test set Dev\_MAN, and Dev\_SEG where English is more prevalent. 
Compared to the SOTA method LPB on the SEAME dataset, CAMEL achieved around a 1\% absolute reduction in MER on both test sets.
\subsubsection{ASRU700+LibriSpeech460 and ASRU200}
Table~\ref{table-asru} shows the results of CAMEL on the ASRU\_Test test set, trained with different datasets, and compares them with several SOTA methods using the same training data.
Compared to the baseline, CAMEL achieves a 7.7\% (8.78\% → 8.10\%) relative reduction in MER using the ASRU700+LibriSpeech460 training data, and a 7\% (11.35\% → 10.55\%) reduction when trained on ASRU200. 
Additionally, when compared with the recent SOTA model SC-MoE, which is trained on ASRU700+LibriSpeech460, CAMEL achieves a 6.9\% (8.66\% → 8.10\%) relative reduction in MER on the ASRU\_Test set.
Moreover, when compared with the LAL~\cite{liu2024aligning} approach trained on ASRU200, CAMEL achieves a reduction of 9.8\% (11.7\% → 10.55\%) in MER.
\begin{table}[t]
    \centering
    \caption{Results on the ASRU200 and ASRU700+LibriSpeech460 datasets, both tested on the ASRU\_Test set, in terms of English WER, Mandarin CER, and Mandarin-English MER (\%):}
    \resizebox{\linewidth}{!}{
        \begin{tabular}{c|cccc|cccc}
            \toprule
            \multirow{2}{*}{System} & \multicolumn{4}{c|}{ASRU700+LibriSpeech460} & \multicolumn{4}{c}{ASRU200}  \\
             \cline{2-5} \cline{6-9} ~ & Params. & MER & CER & WER & Params. & MER & CER & WER \\
            \hline
            CTC-LPB~\cite{liu2024enhancing} & - & - & - & - & - & 11.8 & - & - \\
            LAL~\cite{liu2024aligning} & - & - & - & - & 48.3M & 11.7 & 9.2 & 35.1 \\
            BA-MoE~\cite{chen2023ba} & 43M & 10.28 & 8.16 & 27.48 & - & - & - & - \\
            SC-MoE~\cite{ye2024sc} & 50.2M & 8.66 & 6.50 & 26.36 & - & - & - & -\\
            \midrule
            Baseline & 44.1M & 8.78 & 6.68 & 25.71 & 41.5M & 11.35 & 8.68 & 33.09 \\
            S1 & 44.5M & 8.65 & 6.64 & 25.01 & 41.9M & 11.18 & 8.52 & 32.87 \\
            S2 & 44.5M & 8.60 & 6.56 & 25.25 & 41.9M & 11.10 & 8.42 & 32.92 \\
            S3 & 50.6M & 8.26 & 6.32 & 24.06 & 47.9M & 10.93 & 8.38 & 31.64 \\
            CAMEL & 59.6M & \textbf{8.10} & 6.22 & 23.41 & 56.9M & \textbf{10.55} & 8.04 & 31.00 \\
            \bottomrule
        \end{tabular}
    }
    \label{table-asru}
\end{table}
\subsection{Ablation Study}
To robustly illustrate the effectiveness of our proposed CAMEL system, we conduct ablation experiments on all three datasets.
Specifically, we focus on four key components: the MoE-Adapter with language-wise CTC, the linear gating unit, the gated cross-attention, and the LD decoder.
From the baseline system, systems S1, S2, and S3 are progressively built by adding these modules step by step, culminating into the CAMEL system.
Comparing the S1 system with the Baseline, we can observe the benefits brought by the MoE-Adapter with language-wise CTC. 
For instance, in the system trained with ASRU200, as shown in Table~\ref{table-asru}, adding MoE-Adapter to the end of each latter half encoder layer with language-wise CTC achieves a 1.5\% reduction in MER on the ASRU\_Test dataset (11.35\% → 11.18\%), with similar trends observed on the other two datasets.

The S2 system adds a linear gating unit for a frame-level weighted summation of the Mandarin and English speech representations output by the MoE-Adapter. Although S2 only shows marginal gains across all three datasets, it is essential in constructing the gated cross-attention mechanism. 
As we analyzed, the S3 system, which adds cross-attention before the linear gate to form the gated cross-attention module, shows a notable improvement on each dataset.
S3 reduces the MER on the ASRU\_Test by 4.5\% relative to S1 (8.65\% → 8.26\%) on the ASRU700+LibriSpeech460 dataset.
The above experiments prove the advantage of our proposed gated cross-attention module over a linear gating unit in fusing speech representations of different languages and underscores that cross-lingual contextual information plays a critical role in enhancing code-switching ASR systems.

Last, by comparing the S3 system with CAMEL, we can see the improvements brought by the LD decoder and the attention-based language information bias.
As shown in Table~\ref{table-asru}, CAMEL achieves a 3.5\% reduction in MER on the ASRU\_Test dataset compared to S3 (10.93\% → 10.55\%), trained on ASRU200, with similar trends observed on the other two datasets. 
This shows that the language information bias contributes positively to code-switching ASR.

\section{Conclusion}
This paper proposes the \textbf{C}ross-\textbf{A}ttention enhanced \textbf{M}ixture-of-\textbf{E}xperts (MoE) and \textbf{L}anguage bias (CAMEL) approach for code-switching speech recognition, achieving SOTA performance on the SEAME, ASRU200, and ASRU700+LibriSpeech460 datasets.
Specifically, the CAMEL system demonstrates significant gains from leveraging cross-attention to model cross-lingual contextual information between language-specific speech representations and attention-based language information bias. 
Additionally, we provide detailed ablation studies and analysis of the MoE-Adapter, gated cross-attention, and language diarization decoder, several key modules in CAMEL. 
In the future, we will further explore multi-lingual code-switching ASR and extend CAMEL to scenarios involving three or more languages.


\end{document}